\documentclass[10pt,journal,compsoc]{IEEEtran}
\ifCLASSOPTIONcompsoc
  \usepackage[nocompress]{cite}
\else
  \usepackage{cite}
\fi
\ifCLASSINFOpdf
\else
\fi
\usepackage{amsmath,amsfonts}
\usepackage{algorithmic}
\usepackage{algorithm}
\usepackage{array}
\usepackage{textcomp}
\usepackage{stfloats}
\usepackage{url}
\usepackage{verbatim}
\usepackage{graphicx}
\usepackage{cite}
\usepackage{multirow}
\usepackage{color}
\usepackage{bbm} 
\begin{document}
%
\title{Focusing on Relevant Responses for Multi-modal Rumor Detection}
%
%
%
%

\author{Jun~Li, Yi~Bin,  Liang~Peng,  Yang~Yang,~\IEEEmembership{Senior Member,~IEEE}\\ Yangyang Li, Hao Jin, and Zi Huang 
\IEEEcompsocitemizethanks{
\IEEEcompsocthanksitem Jun Li, Yi Bin, and Liang Peng are with the Center for Future Media and School of Computer Science and Engineering, University of Electronic Science and Technology of China, Chengdu 611731, China. 
\IEEEcompsocthanksitem Yang Yang is with the Center for Future Media, and the School of Computer Science and Engineering, University of Electronic Science and Technology of China, Chengdu 611731, China , and also with the Institute of Electronic and Information Engineering, University of Electronic Science and Technology of China, Guangdong 523808, China.
\IEEEcompsocthanksitem Yangyang Li and Hao Jin are with the National Engineering Research Center for Risk Perception and Prevention, CAEIT, Beijing 100041, China.
\IEEEcompsocthanksitem Zi Huang is with the School of Information Technology and Electrical Engineering, The University of Queensland, Brisbane, Australia. 
\IEEEcompsocthanksitem Corresponding author: Yang Yang. E-mail: dlyyang@gmail.com.}
\thanks{Manuscript received April 19, 2005; revised August 26, 2015.}}

%
%

\markboth{Journal of \LaTeX\ Class Files,~Vol.~14, No.~8, August~2015}%
{Shell \MakeLowercase{\textit{et al.}}: Bare Demo of IEEEtran.cls for Computer Society Journals}
%



\IEEEtitleabstractindextext{%
\begin{abstract}
In the absence of an authoritative statement about a rumor, people may expose the truth behind such rumor through their responses on social media.
Most rumor detection methods aggregate the information of all the responses and have made great progress.
However, due to the different backgrounds of users, the responses have different relevance for discovering the suspicious points hidden in a rumor claim. 
The methods that focus on all the responding tweets would dilute the effect of the critical ones.
Moreover, for a multi-modal rumor claim, the focus of a user may be on several words in the text or an object in the image, so the different modalities should be considered to select the relevant responses and verify the claim.
In this paper, we propose a novel multi-modal rumor detection model, termed \textit{Focal Reasoning Model (FoRM)}, to filter out the irrelevant responses and further conduct fine-grained reasoning with the multi-modal claim and corresponding responses.
Concretely, there are two main components in our FoRM: the coarse-grained selection and the fine-grained reasoning.
The coarse-grained selection component leverages the post-level features of the responses to verify the claim and learns a relevant score of each response.
Based on the relevant scores, the most relevant responses are reserved as the critical ones to the further reasoning.
In the fine-grained reasoning component, we design a relation attention module to explore the fine-grained relations, \textit{i.e.}, token-to-token and token-to-object relations, between the reserved responses and the multi-modal claim for finding out the valuable clues.
Extensive experiments have been conducted on two real-world datasets, and the results demonstrate that our proposed model outperforms all the baselines. 

\end{abstract}

}

\maketitle

\IEEEdisplaynontitleabstractindextext

%
\IEEEpeerreviewmaketitle

\IEEEraisesectionheading{\section{Introduction}\label{sec:introduction}}
\IEEEPARstart{S}{ocial} media has become a popular and important way for people to gather and share information.
A recent survey points out that more than 70\% Americans communicate with others and access news content via social media\footnote{\url{https://www.pewresearch.org/internet/fact-sheet/social-media/}}.
Such hyper-connected social network not only makes information spread faster but also provides an ideal environment for the spread of misinformation~\cite{webb2016digital}.
Among the information in circulation, rumor is the unverified information that could influence public decisions and further lead to social disruption.
For instance, rumors would mislead the voters during the 2016 U.S. election~\cite{allcott2017social} and affect the willingness of the public  to receive the COVID-19 vaccine~\cite{loomba2021measuring}.
Therefore, it is necessary to detect rumors for providing a better network environment and decreasing the detrimental public effects.

\begin{figure}
	\centering		
	\includegraphics[width=\linewidth]{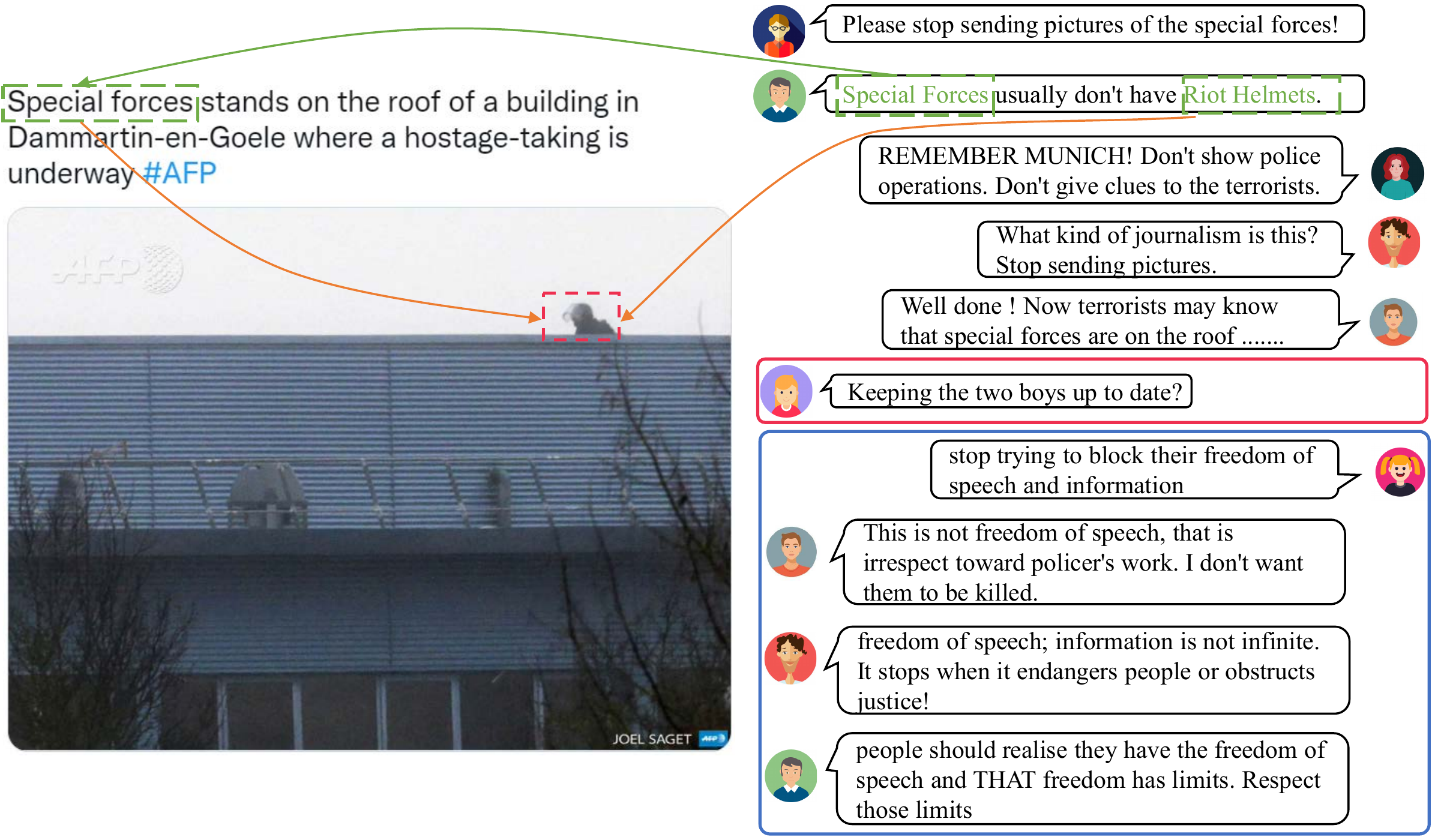}  
	\caption{A conversation constructed with a multi-modal claim and corresponding responses. The tweets in the red box and blue box are irrelevant and unhelpful for verifying the claim respectively. The valuable clues should be captured with the fine-grained reasoning of the responding tweets and the multi-modal claim.} 
	\label{fig:motivation}
\end{figure}

The goal of rumor detection is to verify the truthfulness of a given claim. 
To achieve this goal, fact-checking websites attempt to invite domain experts to confirm the suspected claim.
This manual method is time-consuming and low-coverage, especially on social media, where produces huge amounts of data every day.
In order to automatically verify a claim, some researchers propose to retrieve the relevant evidence from Wikipedia and extract key clues~\cite{thorne2018fact, zhou2019gear, liu2019fine}.
However, some claims on social media, especially breaking news, report newly emerged events that could not be confirmed from existing databases.
These claims always attract public attention and provoke related discussions.
Studies~\cite{procter2013reading, li2015computing}, about the relevant discussion, discovered that the user responses would provide valuable clues for finding out the truth behind an unverified claim.  
With the help of this information, many response-based methods have been proposed and achieved good detection accuracy in rumor detection~\cite{castillo2011information, yang2012automatic, liu2015real, ma2016detecting, chen2018call, zhang2019multi, zhang2021multi, ma2021improving}.

Most response-based methods focus on modelling the textual information.
Early research~\cite{castillo2011information, qazvinian2011rumor, yang2012automatic, liu2015real} takes a lot of efforts on designing hand-crafted features from social content, user characteristics, and information propagation for learning rumor-indicative clues.
However, these hand-crafted features are always statistical information, such as the number of words in a tweet and the fraction of tweets containing a question mark, which lack semantic information of text contents and could not fully represent the complicated rumor-indicative features.
Recently, deep-learning approaches are proposed to leverage the temporal~\cite{ma2016detecting, yu2017convolutional, chen2018call} and structural features~\cite{ma2018rumor,khoo2020interpretable} of the user interactions in a conversation thread.
However, these methods fail to explore multi-modal features of the claim and responding tweets. 
To facilitate the rumor detection process with the multi-modal information, several works~\cite{jin2017multimodal, wang2018eann, zhang2020multimodal} pay more attention on extracting clues from the multi-modal claim.
For example, the att-RNN~\cite{jin2017multimodal} applies the attention mechanism to infer clues from the multi-modal claim and the corresponding social context.
MDDA~\cite{zhang2020multimodal} employs disentangled domain adaption to learn the event-invariant clues hidden in the multi-modal claim. 
Other works~\cite{zhang2019multi, zhang2021multi} fuse the multi-modal claim and relevant responding posts for more complicated clues. 
For instance, Zhang et al.~\cite{zhang2019multi} leverage the multi-modal information for a better post representation and build event memory with the event-invariant features of the sequential responding posts.  

Although the above methods have made great progress in rumor detection, they still suffer from several limitations.
Due to the different backgrounds of users, the responses have different importance in verifying the rumors.
Previous methods that simply leverage all the responding tweets would dilute the effect of the critical ones.  
As the example shown in Fig.~\ref{fig:motivation}, the response in the red box, \textit{``Keeping the two boys up to date?''}, is irrelevant to the multi-modal claim.
In addition, tweets in the blue box are talking about the freedom of speech, which may not provide evidence about the truthfulness of the claim.
These tweets may distract the model from locating the key clues.
Besides, for a multi-modal rumor claim, the focus of a user may be on several words in the text content or an object in the image, so the different modalities should be considered to select the relevant responses and verify the claim. 
Existing multi-modal methods mainly focus on fusing different modalities of the multi-modal claim, but fail to capture the fine-grained (token-level or object-level) interactions between responses and different modalities of the multi-modal claim.
These interactions would connect specific objects in different modalities and further find out the suspicious point to debunk rumors. 
Based on the words \textit{``special forces''} and the place \textit{``on the roof of a building''} in the claim, we locate that the person in the image is a special force.
The response, \textit{``Special Forces usually don't have Riot Helmets''}, connecting to the claim with \textit{``Special Forces''} is inconsistent with the state in the image (the special force wears a riot helmet), which demonstrates the claim is a false rumor.
Without the reasoning of the fine-grained interactions, such suspicious point could not be found for checking the claim. 

Based on the aforementioned considerations, in this paper, we propose a novel multi-modal rumor detection model, termed \textit{Focal Reasoning Model (FoRM)}.
Our FoRM mainly contains two main components: the coarse-grained selection and the fine-grained reasoning.
The aim of the coarse-grained selection component is to filter out the irrelevant response.  
Specifically, we first employ a cross-attention module to fuse the multi-modal information of the claim for a semantic-level representation.
Then, a selection module leverages the post-level features of the responses to verify the multi-modal claim and learns a relevant score of each response.
Based on the relevant scores, the most relevant responses are reserved as the critical ones to the further reasoning.
In the fine-grained reasoning component, we design a relation attention network to construct the full-connected graph of the crucial responses and the multi-modal claim for exploring their fine-grained relations, \textit{i.e.}, token-to-token and token-to-object relations.
Composed of these two components, our model could find out the suspicious point in the multi-modal claim for rumor detection from coarse to fine.
Extensive experiments have been conducted on two real-world datasets, and the results demonstrate that our proposed model outperforms all the baselines. 

The contributions of this paper are as follows:
\begin{itemize}
\item Due to the different backgrounds of users, the responses have different importance in verifying the rumors.
Instead of simply modelling all the responses for rumor detection, \textbf{FoRM} filters out the irrelevant responses, which could further conduct the fine-grained reasoning of the crucial ones and the multi-modal claim.

\item As the focus of a user may be on different modalities of the claim, we design the relation attention network to construct the full-connected graph of the crucial responses and the multi-modal claim for exploring their fine-grained relations, \textit{i.e.}, token-to-token and token-to-object relations.

\item Extensive experiments demonstrate that our \textbf{FoRM} could reach better performance and focus on the  reasonable responses based on the multi-modal claim for detecting rumors.
\end{itemize}

\section{Related Work}
As anyone with an Internet-connected device could share what they may be witnessing or their real-time thoughts on social media~\cite{zubiaga2018detection}, the truthfulness of this information is always uncertain.
Such unverified information, termed the rumor, has the detrimental effect on society and individuals~\cite{zubiaga2016analysing, cao2018automatic}.
Rumor detection aims to identify the veracity of this information and attracts extensive research attention in recent years.
Based on the source of clues which are used to verify rumors, the proposed rumor detection methods could be categorized into three types: the claim-based, the fact-checking, and the response-based.

\textbf{The claim-based methods} attempt to find the clues from the claim, such as the inconsistency~\cite{khattar2019mvae, qian2021hierarchical, qi2021improving} of different modalities or the event-invariant features~\cite{wang2018eann, zhang2020multimodal}.
As attention mechanisms are effective in various multi-modal tasks~\cite{bin2018describing, huang2018csan, liu2020focus, yu2021dual}, Qian et al.~\cite{qian2021hierarchical} design a contextual attention
network to encode the multi-modal context information hierarchically for verifying the truthfulness. 
Instead of utilizing images as text supplements, Qi et al.~\cite{qi2021improving} propose a fine-grained model to capture the inconsistency of the entities in text and image.
Inspired by adversarial network~\cite{goodfellow2014generative, zhao2017learning, yang2018video} and domain adaption~\cite{ma2019deep, li2019joint}, EANN~\cite{wang2018eann} and MDDA~\cite{zhang2020multimodal}  transfer the event-invariant features to newly emerging events.

\textbf{The fact-checking methods} reply on the authoritative sources.
Fact-checking websites, such as politifact.com and snopes.com, invite domain experts to confirm the dubious claim.
To keep up with the enormous volume of generated online information, Ciampaglia et al.~\cite{ciampaglia2015computational} construct a public knowledge graph from Wikipedia for automatically checking claims. 
Recently, deep learning methods are applied to retrieve the evidence from trustworthy sources and infer the fact~\cite{thorne2018fact, zhou2019gear, liu2019fine}. 
However, some claims, especially breaking news, on social media report newly emerged events which could not be checked from the existing database.
For these claims, the fact-checking methods may not work.

\textbf{The response-based methods} actually derive from the crowd wisdom.
Studies~\cite{procter2013reading, li2015computing}, about the relevant discussion, discovered that the community response would provide valuable clues for finding out the truth behind an unverified claim.  
Therefore, how to utilize the abundant information of the community response is a key research direction.
Early studies~\cite{castillo2011information, qazvinian2011rumor, yang2012automatic, liu2015real, wu2015false, ma2015detect} focus on extracting manual features from the contents of messages, user profiles, and diffusion patterns.
For example, Castillo et al.~\cite{castillo2011information} identify four types of hand-crafted features to characterize each topic and train a supervised classifier to debunk rumors.
However, these methods heavily rely on intensive manual efforts and could not fully capture the complicated rumor-indicative features with such statistics.
Then deep-learning methods are proposed to infer clues by modelling the timeline and structure of the user interactions.
Timeline-based methods~\cite{ma2016detecting, yu2017convolutional, chen2018call} concentrate on modelling the temporal pattern of user interactions.
Ma et al.~\cite{ma2016detecting} leverage Recurrent Neural Networks (RNN) to capture the temporal feature of the sequential reply stream. 
To extract the high-level interaction of each part of the reply sequence, the CNN-based framework, CAMI, is proposed in ~\cite{yu2017convolutional}. 
In addition, several works~\cite{ma2018rumor, kumar2019tree, ma2020debunking} argue the user interactions are tree-structured and propose structure-based methods based on Tree-LSTM~\cite{tai2015improved} and Transformer~\cite{vaswani2017attention}.
Specifically, for parallel processing, BCTree LSTM~\cite{kumar2019tree} rebuilds the conversation tree into a binary tree in that each node is always connected with two children. 
To extract better textual and structural features, Tree-Transformer~\cite{ma2020debunking} is proposed and achieves a better performance for conversation trees with different depths. 
Recent studies~\cite{wei2019modeling, khoo2020interpretable, lao2021rumor} attempt to explore both temporal and structural features.
By embedding the time delay and structural information into the multi-head attention layer simultaneously, variants of Transformer~\cite{khoo2020interpretable} are proposed.
In contrast, conversational-GCN~\cite{wei2019modeling} extracts structural and temporal features by GCN and RNN respectively in two steps. 
Lao et al.~\cite{lao2021rumor} design the non-linear structure learning and the linear sequence learning to explore the temporal and the hierarchical characteristic of the user interaction respectively. 
These approaches have shown promising performance on applying deep learning to rumor detection.
However, these methods ignore the multi-modal clues of the thread conversation.
Hence, multimodal fusion methods~\cite{jin2017multimodal, zhang2019multi, zhang2021multi} are proposed to fuse the textual and visual features for detecting rumors.  
The att-RNN~\cite{jin2017multimodal} applies the attention mechanism to explore clues from the multi-modal claim and the corresponding social context.
Zhang et al.~\cite{zhang2019multi} leverage the multi-modal information for a better post representation and build event memory with the event-invariant features of the sequential responding posts.

Instead of considering all the responding tweets in previous works, we propose to select the relevant and valuable tweets.
Furthermore, the fine-grained reasoning is conducted to capture interactions of the multi-modal claim and the selected tweets.

\section{Problem Statement}
In this paper, we employ the textual information of responding tweets to verify the corresponding multi-modal claim.
The conversation thread of a given claim is defined as:
\begin{equation}\label{eq:thread}
	\left\{ \quad
	\begin{aligned}
		&X = \{s, R\}, \\
		&s = \{s_T, s_I\}, \\
        &R = \{r_1, r_2, \cdots, r_N\},
	\end{aligned}
	\right.
\end{equation}
where $s$ is the claim, which consists of the text $s_T$ and the image $s_I$, 
$R$ is the group of the responding tweets which replies to the claim $s$.
Hence, the rumor detection task is to calculate the probability $P(y|X, \theta)$, where $\theta$ is the parameter of rumor classifier and $y$ is the ground truth which belongs to \{\textit{false-rumor, true-rumor, unverified, non-rumor}\}.

\section{Method}
\begin{figure*}
	\centering		
	\includegraphics[width=0.9\linewidth]{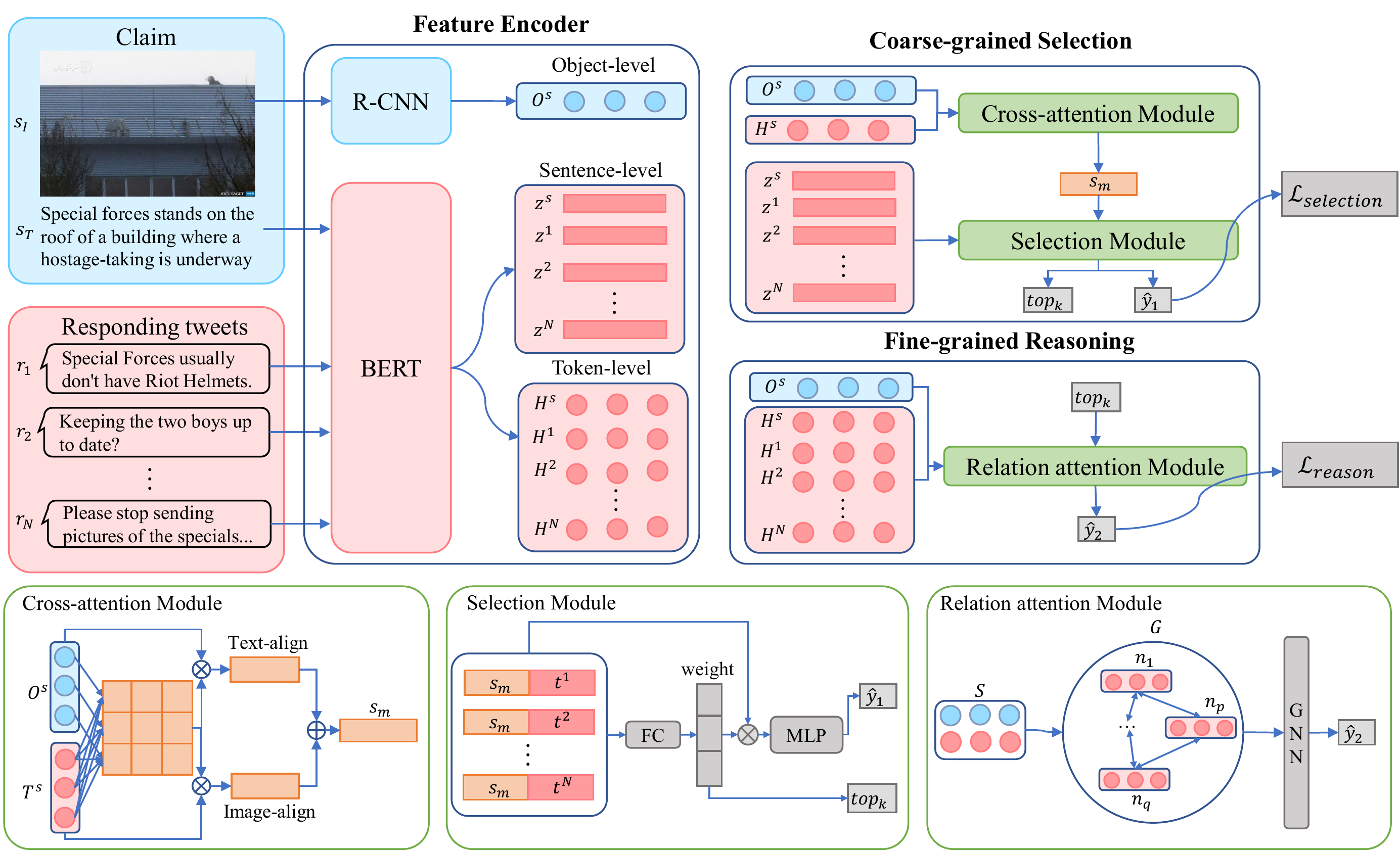}  
	\caption{Our proposed FoRM consists of three main components: feature encoder, coarse-grained selection and fine-grained reasoning. Specifically, the feature encoder extracts the fine-grained (token-level and object-level) features and the coarse-grained (sentence-level) features. The coarse-grained selection applies the cross-attention module to fuse the multi-modal feature of the claim and selects the valuable tweets with their sentence-level features. The fine-grained reasoning leverages the relation attention module to construct a full-connected graph of the selected tweets and the multi-modal claim for capturing their fine-grained relations.}
	\label{fig:framework}
\end{figure*}

In this section, we describe the proposed \textit{Focal Reasoning Model (FoRM)} in detail. 
As illustrated in Fig.~\ref{fig:framework}, our model consists of three main components: feature encoder, coarse-grained selection, and fine-grained reasoning.
To extract the essential features, we employ the textual encoder, BERT~\cite{devlin2018bert}, to extract the token-level and sentence-level features, and the visual encoder, Faster R-CNN model~\cite{anderson2018bottom}, to obtain the object-level features.
The coarse-grained selection component utilizes the cross-attention module to fuse the multi-modal features of the claim and a selection module to learn a relevant score of each response.
Based on the relevant scores, the top k tweets are selected as the critical ones.
The fine-grained reasoning component leverages the relation attention module to model the token-to-token and token-to-object relations between the critical tweets and the claim for finding out the valuable clues.

\subsection{Feature Encoder}
As the model pretrained in a large dataset typically catches the better semantic information, we leverage the pre-trained model as the textual encoder or the visual encoder, to extract the essential features of the text and the image.

\subsubsection{Textual Encoder}
Since BERT has achieved great success in natural language processing tasks~\cite{guan2020deep, li2020survey}, we apply BERT as the textual encoder. 
For coarse-grained selection and fine-grained reasoning, the main purpose of our textual encoder is to extract the coarse-grained (sentence-level) feature and fine-grained (token-level) feature of the text.
We feed the claim text $s_T$ and a responding tweet $r_i$ to BERT separately:
\begin{equation}\label{eq:bert}
    \begin{aligned}
	    &H^s = \text{BERT}(s_T), \\
	    &H^i = \text{BERT}(r_i),
	\end{aligned}
\end{equation}
where $H^s \in \mathbb{R}^{d_t \times M}$ and $H^i \in \mathbb{R}^{d_t \times M}$ are the token hidden states of $s_T$ and $r_i$ respectively, $M$ denotes the number of tokens in  $s_T$ or $r_i$.
We use the hidden state of the first token [CLS] to obtain the representation of the sentence:
\begin{equation}
    \begin{aligned}
	    &z^s = \sigma(W^t H^s_0),\\
	    &z^i = \sigma(W^t H^i_0),
	\end{aligned}
\end{equation}
where $W^t \in \mathbb{R}^{d_t \times d_t}$ is the shared parameter to obtain the sentence-level representations, $\sigma(\cdot)$ is the hyperbolic tangent activation function ($Tanh(\cdot)$), $z^s \in \mathbb{R}^{d_t \times 1}$ and $z^i \in \mathbb{R}^{d_t \times 1}$ denote sentence-level representations of the $s_T$ and $r_i$; 

\subsubsection{Visual Encoder}
VGGNet~\cite{simonyan2014very} and ResNet~\cite{he2016deep} are usually used as the visual encoder in previous work~\cite{zhang2019multi, zhang2021multi, qian2021hierarchical}.
However, they fragment the image into a grid of regions, which barely conform to
the semantics of the image~\cite{peng2020answer}.
To better fuse the semantic information of the text and the image, we apply Faster R-CNN with bottom-up and top-down attention mechanisms as our visual encoder to detect $K$ objects and extract the object-level features of $s_I$:
\begin{equation}\label{eq:fast}
    O^s = \text{Fast-RCNN}(s_I),
\end{equation}
where $O^s=[o_1, o_2, \cdots, o_K] \in \mathbb{R}^{d_i \times K}$ are the features of the $K$ objects in $s_I$. 


\subsection{Coarse-grained Selection}
Social media is open for anyone to respond to any message about a claim.
Due to the different backgrounds of users, the responses have different relevance for discovering the suspicious points hidden in a rumor claim..
To select the critical ones from such tweets, we first apply a cross-attention module to obtain the multi-modal representation of the claim.
Then, we train a selection module to sort all the responding tweets based on their sentence-level features and the multi-modal representation of the claim.

\subsubsection{Cross-attention Module}
Instead of simply concatenating the features of different modalities, we leverage the cross-attention module to align the semantics between them.
Specifically, we first project the token-level feature $H^s$ and the object-level feature $O^s$ into the same dimensional space:
\begin{equation}
    \begin{aligned}
        V^s &= \sigma(W^o O^s), \\
        T^s &= \sigma(W^h H^s),
	\end{aligned}
\end{equation}
where $W^o \in \mathbb{R}^{d \times d_i}$ and $W^h \in \mathbb{R}^{d \times d_t}$ are the training parameters, $V^s=[v_1, v_2, \cdots, v_K] \in \mathbb{R}^{d \times K}$ and $T^s=[w_1, w_2, \cdots, w_M] \in \mathbb{R}^{d \times M}$ are projected object features and token features respectively.
Then, to obtain the relevant objects based on the tokens, we regard each token $w_i \in  T^s$ as the query to calculate the cosine similarity of each object $v_j \in V^s$:
\begin{equation}
    c^{t \rightarrow o}_{ij} = \text{cos}(w_i, v_j).
\end{equation}  
The token-aligned feature is generated by computing the weighted sum of object-level features:
\begin{equation}
    w^{t \rightarrow o}_i = \sum^K_{j=1}c^{t \rightarrow o}_{ij}v_j.
\end{equation}
Finally, we combine all the token-aligned features to the text-aligned representation $s^{t \rightarrow o} \in \mathbb{R}^{d \times 1}$:
\begin{equation}
    s^{t \rightarrow o} = \frac{1}{M}\sum^M_{i=1}w^{t \rightarrow o}_i.
\end{equation}

To obtain the image-aligned representation, we regard the object $v_j$ as the query and all the tokens in the claim text as the key and value:
\begin{equation}
    \begin{aligned}
        c^{o \rightarrow t}_{ji} &= \text{cos}(v_j, w_i), \\
        v^{o \rightarrow t}_j &= \sum^M_{i=1}c^{o \rightarrow t}_{ji}w_j ,\\
        s^{o \rightarrow t} &= \frac{1}{K}\sum^K_{j=1}v^{o \rightarrow t}_j,
	\end{aligned}
\end{equation}
where $c^{o \rightarrow t}_{ji}$ is the cosine similarity of the object $v_j$ and the token $w_i$, $v^{o \rightarrow t}_j$ denotes the object-aligned features of $v_j$, $s^{o \rightarrow t}$ is the image-aligned representation of the claim.

The multi-modal representation of the claim is the weight summation of the text-aligned representation and the image-aligned representation:
\begin{equation}\label{eq:source}
    s_m = \sigma(W^{{o \rightarrow t}}s^{o \rightarrow t}) + \sigma(W^{{t \rightarrow o}}s^{t \rightarrow o}),
\end{equation}
where $W^{o \rightarrow t}$ and $W^{t \rightarrow o}$ are the training parameters, $s_m \in \mathbb{R}^{d \times 1}$ is the multi-modal representation of the claim.

\subsubsection{Selection Module}
In our work, the valuable responding tweet is defined as a tweet that is relevant to the claim and contributes to rumor verification.
To find out the valuable responding tweets, we train a selection module to sort all the responding tweets according to their significance in predicting rumor labels. 

As the claim is the source of the community response, we should provide the context for each responding tweet by combining it with the claim:
\begin{equation}
    A = s_m \mathbbm{1}^T \circ \sigma(W^zZ), 
\end{equation}
where $A \in \mathbb{R}^{2d \times N}$ is the tweet matrix with claim context,  $\mathbbm{1} \in \mathbb{R}^{N \times 1}$ is set to replicate the claim to each responding tweet, $\circ$ is the concatenation operator, $W^z \in \mathbb{R}^{d_t \times d}$ is the parameter.
The volume of the responding tweets is relatively large in real-world scenarios, so we leverage the sentence-level feature of tweets $Z=\{z^1, z^2, \cdots, z^N\} \in \mathbb{R}^{d_t \times N}$ in this component for coarse-grained selection.
Then such tweet matrix is fed to generate the significance of tweets $\alpha \in \mathbb{R}^{1 \times N}$ in predicting rumor label by a \textbf{softmax} function:
\begin{equation}
    \alpha = \textbf{softmax}(W^a A),
\end{equation}
where $W^a \in \mathbb{R}^{1 \times 2d}$ is the training parameter.
To improve the effectiveness of such significance in identifying rumors, we leverage it to predict the rumor label and maximize the accuracy of the prediction: 
\begin{equation}\label{eq:filter}
    \hat{y_1} = \textbf{MLP}(\sum_{i=0}^N\alpha_iz^i),
\end{equation}
where $\alpha_i$ denotes the significance of tweet $z^i$, $\hat{y_1}$ is the coarse-grained prediction, the activation function and the hidden dimension of \textbf{MLP} are ReLU and $d_h$ respectively.
Based on such significance of responding tweets, the top k tweets $R^*=\{r_{1}^*, r_{2}^*, \cdots, r_{k}^*\}$ are selected for further fine-grained reasoning.

\subsection{Fine-grained Reasoning}
The focus of each selected tweet may be an object of the claim image or some tokens in the claim text.
Therefore, we use a relation attention network to model the fine-grained (token-to-token and token-to-object) relations between the selected tweets and the claim. 
Inspired by~\cite{liu2019fine}, the relation attention module constructs a full-connected graph of the selected tweets and multi-modal claim.  
The prediction process is split into two parts: rumor prediction based on a special node $n_q$ after information propagation $P(y|n_q,G,S)$ and significance prediction of such special node $P(n_q|G,S)$:    
\begin{equation}\label{eq:graph-attention}
    P(y|G,S) = \sum_{q \in G} P(y|n_q,G,S)P(n_q|G,S),
\end{equation}
where the node $n_q$ in graph $G$ denotes a selected tweet $r_q \in R^*$ and is initialized with the sentence-level features $z^q$ and token-level features $H^q$, $S$ represents the claim, which consists of the fine-grained feature $S^m=[T^s \circ V^s]$ and the coarse-grained feature $s_m$.

\subsubsection{Information Propagation}
The information propagation aims to aggregate the information of neighbors and generate a new representation.
Hence, we first extract the fine-grained features of neighbor node $n_p$ based on the node $n_q$ and the claim $S$. 
The token-level features $H^p$ of the node $n_p$ is regarded as the query to calculate the cosine similarity of the key ($H^q$ of the node $n_q$ and $S^m$ of the claim):
\begin{equation}
    \begin{aligned}
        T^p, T^q = \sigma(W^p H^p), \sigma(W^q H^q), \\
        C^{p \leftarrow \{q, s\}} = \text{cos}(T^p, [S^m \circ T^q]),
    \end{aligned}
\end{equation}
where $W^p \in \mathbb{R}^{d \times d_t}$ and $W^q \in \mathbb{R}^{d \times d_t}$ are the parameters, $T^p$ and $T^q$ denote the token-level features of $n_p$ and $n_q$ projected into the dimensional space $d$, $C^{p \leftarrow \{q, s\}} \in \mathbb{R}^{M \times (2M+K)}$ is the cosine similarity matrix.  
Then, we obtain fine-grained features $T^{p \leftarrow \{q, s\}}$ of the neighbor $n_p$:  
\begin{equation}
    \begin{aligned}
        \alpha^{p \leftarrow \{q, s\}} &= \textbf{softmax}(C^{p \leftarrow \{q, s\}}), \\
        T^{p \leftarrow \{q, s\}} &= \alpha^{p \leftarrow \{q, S\}}[S \circ T^q] + T^p,
    \end{aligned}
\end{equation}
where $\alpha^{p \leftarrow \{q, s\}} \in \mathbb{R}^{M \times (2M+K)}$ denotes the significance of the tokens in $n_p$ and the tokens and objects in $S$.
To the end, we fuse all the token features in $T^{p \leftarrow \{q, s\}}$ to obtain the representation of $n_p$ with the attention mechanism:
\begin{equation}
    \begin{aligned}
        &\beta = \textbf{softmax}(W^{p \leftarrow \{q, s\}}T^{p \leftarrow \{q, s\}})), \\
        &z^{p \leftarrow \{q, s\}} = \sum^M_{i=1}\beta_i T^{p \leftarrow \{q, s\}}_i,
    \end{aligned}
\end{equation}
where $\beta \in \mathbb{R}^{1 \times M}$ denotes the weight of all the token in $n_p$, $W^{p \leftarrow \{q, s\}} \mathbb{R}^{1 \times d}$ is the training parameter, $z^{p \leftarrow \{q, s\}} \in \mathbb{R}^{1 \times d}$ is the representation of $n_p$,.

To generate the propagated representation of $n_q$, we aggregate the representations of all its neighbors.
We fuse the neighbor's representation $z^{p \leftarrow \{q, s\}}$, the sentence-level features $z^q$ of $n_q$ and the multi-modal feature $s_m$ of the claim to calculate the importance of $\lambda_p$ the neighbor node $n_p$: 
\begin{equation}
    \lambda_p= \textbf{softmax}_p(\textbf{MLP}(z^{p \leftarrow \{q, s\}} \circ s_m \circ \sigma(W^zz^q)),
\end{equation}
where $\textbf{softmax}_p$ selects the value of $n_p$ from the output of \textbf{softmax} function.
The propagated representation $v^q$ of $n_q$ is represented as:
\begin{equation}\label{eq:node-representation}
    v^q = (\sum_{p \in G}\lambda_p z^{p \leftarrow \{q, s\}}) \circ \sigma(W^zz^q).
\end{equation}
Finally, we predict the probability of the rumor label based on the special node $n_q$:
\begin{equation}\label{eq:node-prediction}
    P(y|n_q,G,S) = \textbf{softmax}(W^y(v^q \circ s_m)),
\end{equation}
where $W^y \in \mathbb{R}^{4 \times 3d}$ is the parameter.

\subsubsection{Significance Prediction}
Each node represents a responding tweet which may provide clues to verify the claim, so we use the claim to evaluate the significance of a node. 

Given a node $n_q$, we calculate the similarity between the fine-grained claim features $S^m$ and tokens in $T^q$:
\begin{equation}\label{eq:node-cos}
    C^{s \leftarrow q} = \text{cos}(S^m, T^q),
\end{equation}
where $C^{s \leftarrow q} \in \mathbb{R}^{(M+K) \times M}$ is the translation matrix between the multi-modal claim and the tokens in $T^q$.
Then, we obtain the claim fine-grained features $S^{s \leftarrow q} \in \mathbb{R}^{d \times (M+K)}$ by the weighted summation of all the tokens' features in $T^q$:  
\begin{equation}
    \begin{aligned}
        \alpha^{s \leftarrow q} &= \textbf{softmax}(C^{s \leftarrow q}), \\
        S^{s \leftarrow q} &= \alpha^{s \leftarrow q} T^q + S^m.
    \end{aligned}
\end{equation}
The significance of $n_q$ is predicted as:
\begin{equation}
    \begin{aligned}
        s^{s \leftarrow q} &= \sum_{i=1}^{M+K}S^{s \leftarrow q}_i, \\
        P(n_q|G,S) &= \textbf{softmax}_q(W^{s \leftarrow q}s^{s \leftarrow q}),
    \end{aligned}
\end{equation}
where $s^{s \leftarrow q} \in \mathbb{R}^{d \times 1}$ is the representation vector of the claim, $W^{s \leftarrow q} \in \mathbb{R}^{1 \times d}$ is the parameter, $\textbf{softmax}_q$ selects the value of $n_q$ from the output of \textbf{softmax} function.
Finally, the rumor prediction by the whole graph is aggregated following Eq.~\ref{eq:graph-attention}.

\subsection{Model Training}
Our model contains two components to identify rumors: the coarse-grained selection and the fine-grained reasoning.
The coarse-grained selection is used to find out valuable responding tweets by generating the significance of each tweet.
Hence, to improve the effectiveness of such significance in identifying rumors, we design the selection loss to maximize the probability of the prediction $y$ in the coarse-grained selection: 
\begin{equation}\label{eq:filter-loss}
    \mathcal{L}_{selection} = \textbf{CrossEntropy}(y, \hat{y_1}),
\end{equation}
where $y$ denotes the ground truth. 

In fine-grained reasoning, the model predicts the rumor label based on the token-to-token and token-to-object relations of the selected tweets and the claim, thus we leverage the cross entropy loss to train our model, which minimizes the difference between the ground truth $y$ and the predicted distribution $ P(y|G,S)$:
\begin{equation}\label{eq:classifer-loss}
    \mathcal{L}_{reason} = \textbf{CrossEntropy}(y, P(y|G,S)).
\end{equation}

Towards the end, the total loss function is defined as the linear combination of the above two loss functions:
\begin{equation}\label{eq:total-loss}
    \mathcal{L} = \mathcal{L}_{selection} + \mathcal{L}_{reason}.
\end{equation}

\section{experiments}
\subsection{Experimental Setting}
\subsubsection{Datasets}
We train and evaluate our FoRM on two public Twitter datasets: Twitter15 and Twitter16~\cite{ma2017detect}.
Both datasets consist of Twitter threads, which contain a claim and a set of responding tweets.  
To expand a given claim into a multi-modal one, we crawl the corresponding image based on its ID and textual information. 
For the responding tweets, we observe that retweets in both datasets just repeat the textual content of the claim which could not provide clues for our model.
Following~\cite{khoo2020interpretable}, we remove all the retweets and conduct the 5-fold cross-validation.
The statistics of the processed datasets are shown in Table~\ref{tab:data} and the processed data is publicly accessible\footnote{\url{https://github.com/serenaklm/rumor_detection}}.
\begin{table}[]
    \renewcommand\arraystretch{1.2}
	\centering
	\caption{The statistics of the datasets} 
	\label{tab:data}
	\begin{tabular}{c c c} \hline
		Statistic & TWitter15 & Twitter16 \\ 
		\hline
		Total conversation & 1413 & 756\\ 
		Total tweets & 57,368 & 27,652\\ 
		Unverified & 358 & 190\\ 
		True & 350 & 189\\ 
		False & 334 & 172\\ 
        Non-rumor & 371 & 205\\
		images & 737 & 400\\
		\hline
	\end{tabular}
\end{table}

\subsubsection{Implementation Details}
In our experiments, the textual encoder inherits huggingface’s implementation\footnote{\url{https://github.com/huggingface/transformers}} and the visual encoder is implemented by bottom-up-attention\footnote{\url{https://github.com/peteanderson80/bottom-up-attention}}.
Limited by the memory of GPUs, we set the number of responding tweets $N$ to 100, the max number of tokens $M$ to 35, and the max number of objects $K$ to 36.
We use the special token [PAD] in BERT embedding to fill the tweets less than 100 and the tokens less than 35.
For the objects less than 36, we leverage the one-padding to fill them.
The dimension of the textual (token-level, sentence-level) feature $d_t$ and object-level feature $d_i$ are 768 and 2048 respectively. 
To fuse features of different modalities, we project them into the same dimensional space $d=768$. 
To further capture the fine-grained clues, we select $top_5$ tweets and $top_{10}$ tweets for Twitter15 and Twitter16 in coarse-grained selection respectively.
The hidden dimension $d_h$ of \textbf{MLP} is set to 128.
Finally, the model is optimized by Adam optimizer. 
We set the learning rate and batch size to 5e-5 and 4 respectively.

\begin{table*}
    \renewcommand\arraystretch{1.2}
	\centering
	\caption{Rumor verification results of two datasets (F: False rumor; T: True rumor; U: Unverified rumor; NR: Non-rumor).}
	\label{tab:res_p}
	\begin{tabular}{c |c |ccccc|ccccc}
        \hline
         & \multirow{2}{*}{\textbf{Method}} & \multicolumn{5}{c}{\textbf{Twitter15}} & \multicolumn{5}{c}{\textbf{Twitter16}} \\
        \cline{3-12}
          &&\textbf{Accuracy} & \textbf{F} & \textbf{T} & \textbf{U} & \textbf{NR} & \textbf{Accuracy} &  \textbf{F} & \textbf{T} & \textbf{U} & \textbf{NR}\\
        \hline
        \multirow{5}{*}{\textbf{Textual}}
        & DTC & 0.454 & 0.355 & 0.317 & 0.415 & 0.733 & 0.465 &  0.393 & 0.419 & 0.403 & 0.643 \\
        & GRU2 & 0.641 & 0.634 & 0.688 &  0.571 & 0.684 & 0.633 &  0.715 &  0.577 & 0.527 & 0.617 \\
        & CAMI & 0.653 & 0.603 & 0.686 &  0.505 & 0.719 & 0.691 &  0.597 &  0.767 & 0.658 & 0.739 \\
        & RvNN & 0.705 & 0.710 & 0.721 &  0.730 & 0.655 & 0.806 &  0.755 &  0.893 & 0.830 & 0.734 \\
        & PLAN & 0.845 & 0.858 & 0.895 &  0.802 & 0.823 & 0.874 &  0.839 &  0.917 & \textbf{0.888} & 0.853 \\
        \hline
        \multirow{3}{*}{\textbf{Multi-modal}}
        & EANN & 0.682 & 0.701 & 0.792 &  0.660 & 0.561 & 0.758 &  0.701 &  0.848 & 0.84 & 0.648 \\
        & att-RNN & 0.695 & 0.660 & 0.818 &  0.626 & 0.647 & 0.737 &  0.709 &  0.854 & 0.746 & 0.631 \\
        & MKEMN & 0.807 & 0.791 & 0.870 &  0.704 & 0.865 & 0.817 &  0.684 &  0.892 & 0.836 & 0.860 \\
        \hline
        & \textbf{FoRM} & \textbf{0.877} & \textbf{0.884} & \textbf{0.896} &  \textbf{0.840} & \textbf{0.869} & \textbf{0.886} &  \textbf{0.853} &  \textbf{0.940} & 0.880 & \textbf{0.866} \\
        \hline
    \end{tabular}
\end{table*}
\subsection{Baselines}
We compare the proposed FoRM with the following two categories of baselines:

\subsubsection{Single Modality Methods}
\begin{itemize}
\item \textbf{DTC~\cite{castillo2011information}:} The early method identifies four types of hand-crafted features including message-based features, user-based features, topic-based features, and propagation-based features. By training a decision tree as the classifier, DTC predicts the credibility of a Twitter event.

\item \textbf{GRU2~\cite{ma2016detecting}:} The deep learning method regards the responding tweets as a chronological sequence and constructs the variable-length time series to represent the dense popularity of the claim in the diffusion. The multilayer GRU network is leveraged to extract such temporal pattern of the relevant tweets.

\item \textbf{CAMI~\cite{yu2017convolutional}:} Considering the long-tailed distribution, CAMI normalizes the timestamp and slices the responding tweets into equal groups in chronological order. By aggregating the representation of each group, this method captures the local and global features of the rumor event.

\item \textbf{RvNN~\cite{ma2018rumor}:} To model the non-sequential
propagation structure of responding tweets, a thread is organized as the tree-structure propagation based on the reply relationship. RvNN recursively generates the representation of such propagation from different angles (bottom-up or top-down) to identify rumors. 

\item \textbf{PLAN~\cite{khoo2020interpretable}:} Arguing that the structure in a thread is not explicit, a Transformer-based method, PLAN, is proposed to model the full-connected structure, \textit{i.e.}, the user could reply to any tweets in the thread.
\end{itemize}

\subsubsection{Multi-modal Methods}
\begin{itemize}
\item \textbf{EANN~\cite{wang2018eann}:} To debunk fake news in newly emerged events, EANN applies the adversarial neural networks to capture event-invariant features of the multi-modal claim.

\item \textbf{att-RNN~\cite{jin2017multimodal}:} To detect rumors, att-RNN leverages the attention mechanism to extract the multi-modal features from the claim, including textual, visual and social context features.

\item \textbf{MKEMN~\cite{zhang2019multi}:} MKEMN learns the better representations for responding tweets by fusing the textual, visual and background knowledge features. To transfer the event-invariant features from seen data to newly emerged events, MKEMN models all the responding tweets chronologically to extract the current event's latent information and builds the external memory to capture the event-invariant ones.
\end{itemize}

\begin{figure}
	\centering		
	\includegraphics[width=\linewidth]{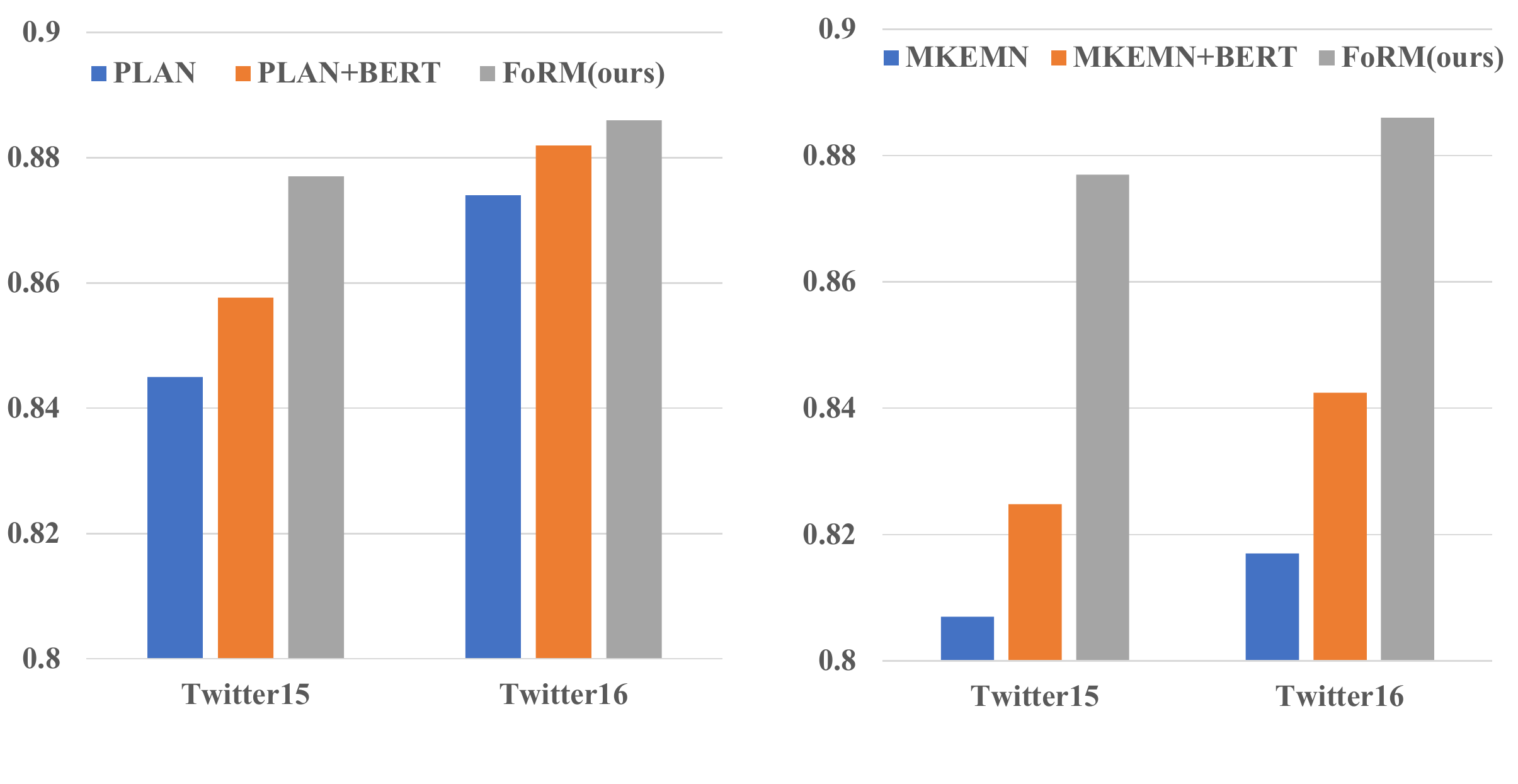}  
	\caption{Performance comparison with PLAN+BERT and MKEMN+BERT.} 
	\label{fig:PLAN+MKEMN}
\end{figure}

\begin{table*}
    \renewcommand\arraystretch{1.2}
	\centering
	\caption{The effectiveness of different components of FoRM.}
	\label{tab:ablation}
	\begin{tabular}{l |ccccc|ccccc}
        \hline
        \multirow{2}{*}{\textbf{Method}} & \multicolumn{5}{c}{\textbf{Twitter15}} & \multicolumn{5}{c}{\textbf{Twitter16}}\\
        \cline{2-11}
          &\textbf{Accuracy} & \textbf{F} & \textbf{T} & \textbf{U} & \textbf{NR} &\textbf{Accuracy} & \textbf{F} & \textbf{T} & \textbf{U} & \textbf{NR}\\
        \hline
        FoRM w/o V & 0.850 & 0.862 & 0.904 & 0.815 & 0.823 & 0.854 & 0.822 & 0.909 & 0.866 & 0.825 \\
        FoRM w/o F & 0.806 & 0.793 & 0.869 & 0.744 & 0.808 & 0.843 & 0.797 & 0.896 & 0.884 & 0.800  \\
        FoRM w/o S & 0.835 & 0.832 & 0.865 & 0.817 & 0.821 & 0.853 & 0.813 & 0.912 & 0.867 & 0.822 \\
        \textbf{FoRM} & 0.877 & 0.884 & 0.896 &  0.840 & 0.869 & 0.886 &  0.853 &  0.940 & 0.880 & 0.866 \\
        \hline
    \end{tabular}
\end{table*}

\begin{table}
    \renewcommand\arraystretch{1.2}
	\centering
	\caption{The effectiveness of visual encoder on Twitter15.}
	\label{tab:ablation_rcnn15}
	\begin{tabular}{l |ccccc}
        \hline
        \multirow{2}{*}{\textbf{Method}} & \multicolumn{5}{c}{\textbf{Twitter15}} \\
        \cline{2-6}
          &\textbf{Accuracy} & \textbf{F} & \textbf{T} & \textbf{U} & \textbf{NR} \\
        \hline
        FoRM+ResNet & 0.859 & 0.859 & 0.909 & 0.838 & 0.832 \\
        FoRM+VGG & 0.851 & 0.864 & 0.898 & 0.815 & 0.828  \\
        \textbf{FoRM} & 0.877 & 0.884 & 0.896 &  0.840 & 0.869  \\
        \hline
    \end{tabular}
\end{table}

\begin{table}
    \renewcommand\arraystretch{1.2}
	\centering
	\caption{The effectiveness of visual encoder on Twitter16.}
	\label{tab:ablation_rcnn16}
	\begin{tabular}{l |ccccc}
        \hline
        \multirow{2}{*}{\textbf{Method}} & \multicolumn{5}{c}{\textbf{Twitter16}} \\
        \cline{2-6}
          &\textbf{Accuracy} & \textbf{F} & \textbf{T} & \textbf{U} & \textbf{NR} \\
        \hline
        FoRM+ResNet & 0.860 & 0.828 & 0.905 & 0.866 & 0.838 \\
        FoRM+VGG & 0.858 & 0.828 & 0.912 & 0.860 & 0.832  \\
        \textbf{FoRM} & 0.886 &  0.853 &  0.940 & 0.880 & 0.866  \\
        \hline
    \end{tabular}
\end{table}
\subsection{Performance Comparison}
Table~\ref{tab:res_p} demonstrates the rumor verification results of the baselines and our model on both datasets.
Following~\cite{ma2018rumor}, we leverage the accuracy as the evaluation metric over the four classes and $F_1$ score for each one.
Compared with all the baselines, our model, FoRM, reaches the better performance, which indicates the advantages of exploring the fine-grained multi-modal clues from the community response.  

The single modality methods model community response from different perspectives.
Without the deep learning, DTC performs worse than other methods, because the hand-crafted features could not enumerate all the features hidden in the community response.
Although the deep learning network is applied in the other models, the results of GRU2 and CAMI are poor due to their failure to capture the comment of a special user and the interaction between each tweet. 
RvNN and PLAN first obtain the representation of each responding tweet and model the interaction between them based on their representation, so they reach the better performance. 
RvNN captures the recursive relations in a tree structure, which means that the tweet only replies to its father tweet. 
In contrast, PLAN argues that the interaction is unclear and constructs a full-connected graph, which indicates that each tweet may reply to every tweet in the thread, \textit{e.g.}, a tweet denies its father tweet and provides the evidence to support the claim.
Therefore, RvNN may extract more complicated interaction features to achieve the best result among all the single modality methods.

To model the multi-modal information, EANN and att-RNN introduce the visual information to the claim and rely on the multi-modal claim to identify rumors.
Their results are even better than those of GRU2 and CAMI.
The reason may be that the writing style of the claim is different among various rumours. 
Furthermore, combining the attached image would provide the multi-model clues for learning the style of different claims.
This indicates that the claim, especially the multi-modal claim, may be more important to detect rumors.
Compared with EANN and att-RNN, MKEMN considers the responding tweets in a thread and improves the performance.
However, MKEMN utilizes the multi-modal information to obtain a better post representation without modelling the interaction between different tweets.
This may be the reason of its poorer performance compared to PLAN.

To the end, our FoRM not only leverages the multi-modal information of the claim, but also considers the interaction between the responding tweets.
Specifically, we assume that a part of responding tweets may be irrelevant and filer them out, rather than modelling the interaction across all tweets.
For the reserved tweets, instead of extracting the multi-model clues in post-level, we propose a fine-grained method to capture the multi-model clues in token-to-token and token-to-object level.
Hence, FoRM achieves the best performance among not only single modality methods but also the multi-modal methods.

Furthermore, we use the pre-trained BERT as the textual encoder of PLAN and MKEMN for fair comparison.
These variants are denoted with PLAN+BERT and MKEMN+BERT in Fig.~\ref{fig:PLAN+MKEMN} respectively. 
As we can see, the pre-trained BERT improves the performance of PLAN and MKEMN on both datasets.
This suggests that features extracted from the pre-trained model are effective to capture more complicated clues and achieve better accuracy.
However, different from PLAN and MKEMN which simply leverage such features for a better tweet representation, our model applies the relation attention network to capture the fine-grained relations of different tokens.
As shown in Fig.~\ref{fig:PLAN+MKEMN}, FoRM still outperforms PLAN+BERT and MKEMN+BERT.

\subsection{Ablation Study}


To validate the effectiveness of the components in FoRM, we design some ablation experiments on both datasets. 

First, we remove the key component of our model and design three variants of FoRM as follows:
\begin{itemize}
\item \textbf{FoRM w/o V:} The image in a claim could help the model to find out multi-model clues for identifying rumors.
To evaluate the effectiveness of the image, this variant of FoRM removes the visual information of the claim and only captures the textual clues.

\item \textbf{FoRM w/o F:} To model the fine-grained relations, the relation attention module connects different tokens and objects between responding tweets and the claim. 
In this variant of FoRM, we remove the relation attention module and model all the responding tweets without the fine-grained reasoning.

\item \textbf{FoRM w/o S:} The selection loss function is used to improve the effectiveness of the significance about each responding tweet in the coarse-grained selection, which could further affect the selected top k tweets.
To explore the effect of such loss function, this variant removes the selection loss function.
\end{itemize}
\begin{figure*}
	\centering		
	\includegraphics[width=\linewidth]{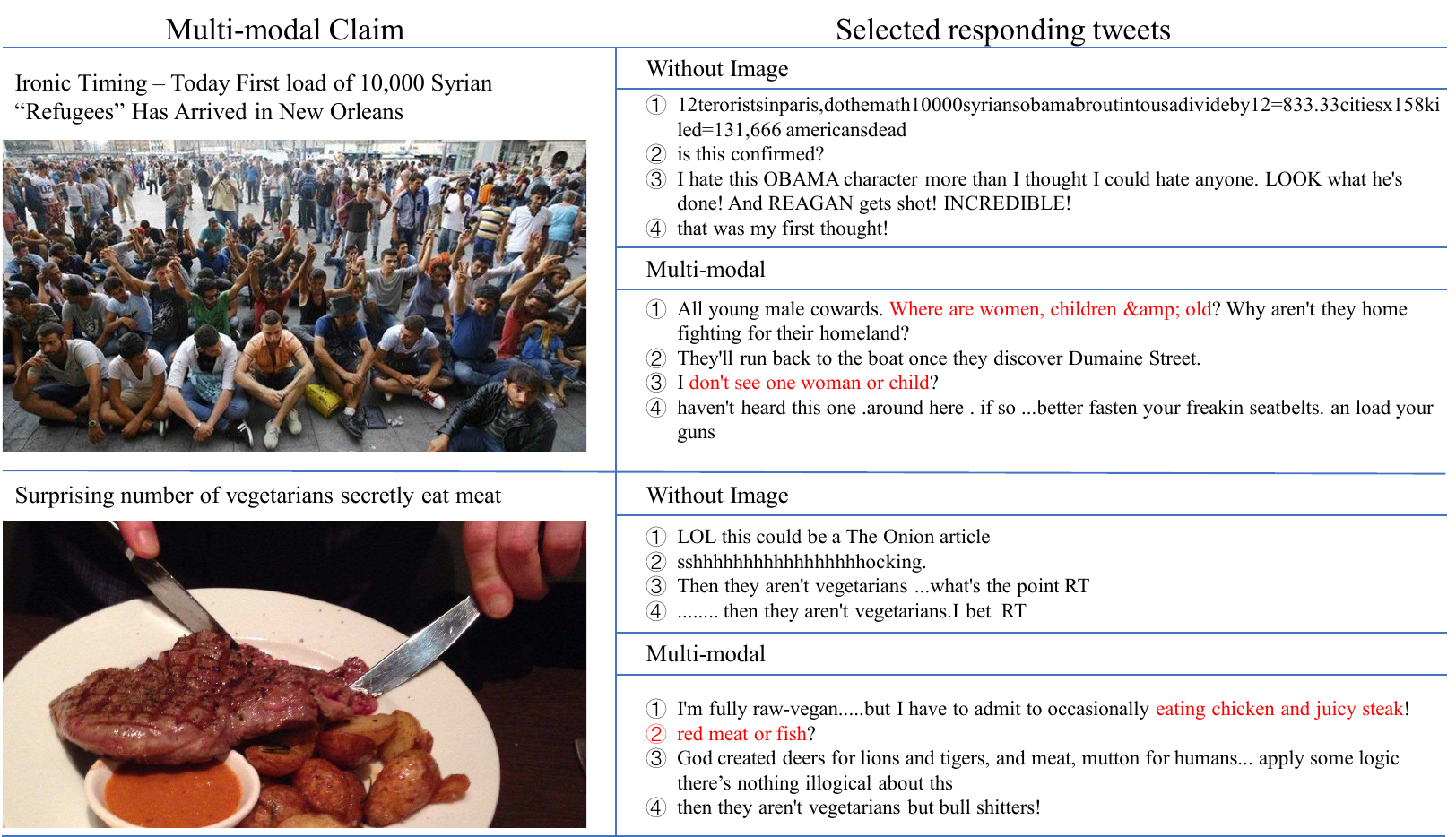}  
	\caption{Samples of multi-modal claims. For each claim, we select and sort the responding tweets with or without the visual information. In the end, we show the top four tweets from the selected results. The red words indicate the clues explored from the responding tweets.} 
	\label{fig:case}
\end{figure*}

As shown in Table~\ref{tab:ablation}, without the fine-grained reasoning, the accuracy of FoRM w/o F drops significantly.
The reason may be: 
(1) simply concatenation of the claim and responding tweets in post-level that could not capture fine-grained clues;
(2) aggregation of all the responding tweets would dilute the effect of the crucial ones.
Furthermore, we observe that without the selection loss function, the selection module could not be supervised to select the valuable tweets for further fusion.
As not all the threads contain multi-modal claims, we generate a dummy image for claims without visual information, which may result in less performance degradation of FoRM w/o V. 

Second, we compare the performance of FoRM with different visual encoders. 
As the pre-trained VGG19 and ResNet101 are widely used for extracting the visual feature, we leverage them to replace our visual encoder respectively.
As illustrated in Table~\ref{tab:ablation_rcnn15} and Table~\ref{tab:ablation_rcnn16},  
compared with FoRM, the accuracy of FoRM+ResNet and FoRM+VGG drop about 2\% and 3\% on Twitter15 and Twitter16 respectively.
These results indicate that the feature extraction of Faster R-CNN conforms to the semantics of the image, which performs better in our model.

\begin{figure}[htbp]
	\centering		
	\includegraphics[width=\linewidth]{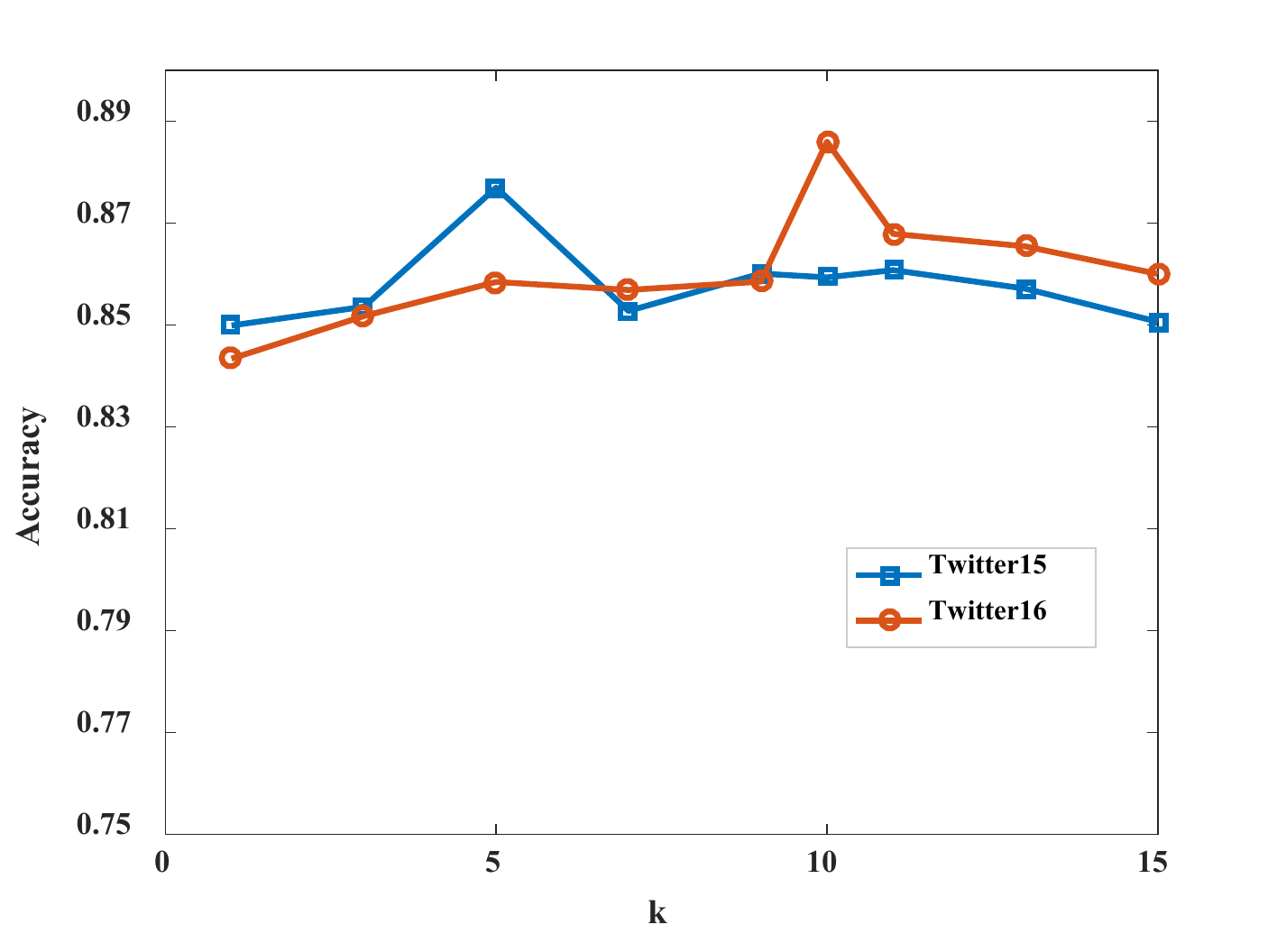}  
	\caption{The influence of the number k of selected tweets on the performance of our model.} 
	\label{fig:topk}
\end{figure}
Finally, Fig.~\ref{fig:topk} shows the influence of the number of selected tweets ($top_k$) on performance.
We observe that the accuracy of our model increases from 1 to 5 while it decreases after k=5 on Twitter15.
As for Twitter16, the accuracy increases from 1 to 10 and decreases after k=10.
Hence, we set k=5 and k=10 to achieve the best performance for Twitter15 and Twitter16 respectively.
The results also suggest that the more responding tweets used, the worse performances may be obtained, since more tweets may provide more noise in rumor identification. 


\subsection{Case Study}
Fig.~\ref{fig:case} shows the examples of multi-modal claims and their selected responding tweets.
We compare the selected responding tweets with and without the visual information.
In upper false rumor, without considering the attached image, the selected tweets focus on the text of the claim and show the protest against the arrival of refugees.
In contrast, our multi-modal method, FoRM, combines the image and the text in the claim and finds the vital clue from the responding tweets that the refugees in the picture are all young males.
Besides, the selected tweets further point out that there are no women, children, or the old in the picture which is inconsistent with common knowledge related to refugees.
Hence, our model doubts the authenticity of this claim and identifies it as a false rumor. 
The bottom example is an unverified rumor, which indicates the claim is still under discussion and its truth value remains unresolved~\cite{zubiaga2018detection}.
Since the image and the text in the claim do not include suspicious points to identify its truthfulness, the methods with and without the image both select subjective tweets which express the feeling or the experience of a user.
To be more specific, the selected tweets of the textual method are about the feelings of users, which are anger or shock at vegetarians eating meat secretly.  
As for our method, the selected tweet is about someone's experiment which is not only relevant to vegetarians in the text but also about the steak in the image.
However, because the image does not include vegetarians, FoRM may more concentrate on the object ``steak'' in the image and the word ``meat'' in the text and select the unimportant tweet, ``red meat or fish'', which is not relevant to the topic (vegetarians) of the claim.



\section{CONCLUSION}
In this paper, we proposed a multi-modal rumor detection model, termed FoRM, to filter out the irrelevant responses and further conduct fine-grained reasoning with the multi-modal claim and corresponding responses.
Specifically, the volume of the responding tweets was relatively large in real-world scenarios, so we designed a coarse-grained selection module to leverage the post-level features for verifying the claim and learns the relevant score of each response.
Based on the relevant scores, the top k tweets were selected as the critical ones. 
We then applied a relation attention module to capture the fine-grained relations, \textit{i.e.}, token-to-token and token-to-object relations, of the multi-modal claim and the critical tweets for debunking rumors.
The experiments on two public datasets showed that our model outperforms all the baselines and could select more reasonable tweets based on the multi-modal claim for further rumor detection.

\bibliographystyle{IEEEtran}
\bibliography{ref}

\begin{IEEEbiographynophoto}{Jun Li} received the bachelor’s degree and master’s degree from Chongqing University in 2013 and 2016. He is currently pursuing the Ph.D. degree with the School of Computer Science and Engineering, University of Electronic Science and Technology of China, Chengdu, China. His current research interests include Natural Language Processing and Social Media Analytics.
\end{IEEEbiographynophoto}
\begin{IEEEbiographynophoto}{Yi Bin} is currently with the University of Electronic Science and Technology of China, Chengdu, China. He received the Ph.D. degree from UESTC in 2020. His research interests include multimedia analysis, vision understanding and deep learning.
\end{IEEEbiographynophoto}
\begin{IEEEbiographynophoto}{Liang Peng} received his Ph.D. degree in School of Computer Science and Engineering, University of Electronic Science and Technology of China. His research interests include vision understanding, natural language processing and deep learning.
\end{IEEEbiographynophoto}
\begin{IEEEbiographynophoto}{Yang Yang} \textbf{(Senior Member, IEEE)} received the bachelor’s degree from Jilin University, Changchun, China, in 2006, the master’s degree from Peking University, Beijing, China, in 2009, and the Ph.D. degree from The University of Queensland, Brisbane, QLD, Australia, in 2012, all in computer science. He is currently with the University of Electronic Science and Technology of China, Chengdu, China. His current research interests include multimedia content analysis, computer vision, and social media analytics.
\end{IEEEbiographynophoto}
\begin{IEEEbiographynophoto}{Yangyang Li} received the B.S. degree from the Nanjing University of Information and Technology, in 2009, and the Ph.D. degree in computer science from the Beijing University of Posts and Telecommunications, in 2015. He is currently a Senior Engineer at the National Engineering Research Center for Risk Perception and Prevention. His research interests include the cyberspace security, social networks, and data science.
\end{IEEEbiographynophoto}
\begin{IEEEbiographynophoto}{Hao Jin} received the B.S. degree in computer science from the Hefei University of Technology in 2013, and the Ph.D. degree in Communication and Information System from the University of Chinese Academy of Sciences, in 2018. She is currently a Senior Engineer at the National Engineering Research Center for Risk Perception and Prevention. Her research interests include the cyberspace security and social networks.
\end{IEEEbiographynophoto}
\begin{IEEEbiographynophoto}{Zi Huang} received her BSc degree from the Department of Computer Science, Tsinghua University, China, and the PhD degree in computer science from The University of Queensland. She is a professor and ARC Future Fellow in The School of Information Technology and Electrical Engineering, The University of Queensland. Her research interests mainly include multimedia indexing and search, social data analysis, and knowledge discovery.
\end{IEEEbiographynophoto}

\end{document}